%% file: tester.tex
\documentclass[a4paper,11pt]{article}
\pdfoutput=1 %

\usepackage{jinstpub} %

\usepackage{lineno}
\input{wz_commands}

\title{\boldmath SMX and front-end board tester for CBM readout chain}

\author[a]{Wojciech M. Zabołotny,\note{Corresponding author.}}
\author[b]{David Emschermann,}
\author[a]{Marek Gumiński,}
\author[a]{Michał Kruszewski,}
\author[b]{Jörg Lehnert,}
\author[a]{Piotr Miedzik,}
\author[a]{Krzysztof Poźniak,}
\author[a]{Ryszard Romaniuk,}
\author[b]{Christian J. Schmidt}

\affiliation[a]{Institute of Electronic Systems, Warsaw University of Technology,\\
             Nowowiejska 15/19, 00-665 Warszawa, Poland}
\affiliation[b]{GSI - Helmholtzzentrum für Schwerionenforschung GmbH,\\
             Darmstadt, Germany}

\emailAdd{wzab@ise.pw.edu.pl}

\abstract{
The \textcolor{black}{STS-MUCH-XYTER (SMX)} chip is a front-end ASIC dedicated to the readout of 
\textcolor{black}{Silicon Tracking System (STS)} and \textcolor{black}{Muon Chamber (MUCH)} detectors in 
\textcolor{black}{the Compressed Baryonic Matter (CBM) experiment}.
The production of the ASIC and the front-end boards based on it is just being started and requires thorough testing to assure quality.
The paper describes the SMX tester based on a standard commercial Artix-7 FPGA module with an additional simple baseboard.
In the standalone configuration, the tester is controlled via IPbus and enables full functional testing of connected SMX, \textcolor{black}{front-end board (FEB)}, or a full detector module.
The software written in Python may easily be integrated with higher-level testing software.
}


\arxivnumber{2110.10619} %

\proceeding{TWEPP 2021\\
  20--24.09.2021}

\begin{document}
\maketitle
\flushbottom

\section{Introduction}
\label{sec:intro}
\begin{figure}[htbp]
	\centering %
	\includegraphics[width=.55\linewidth]{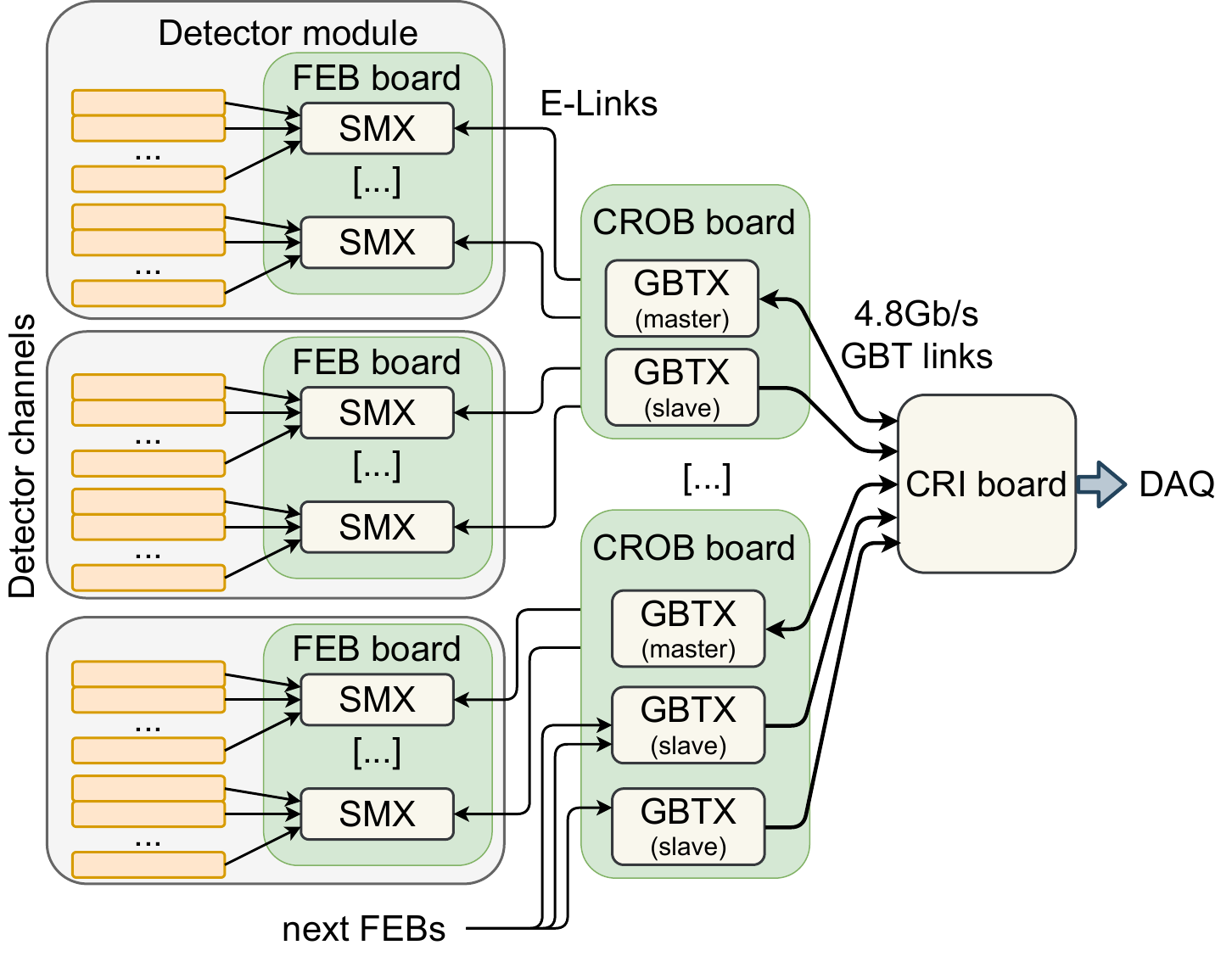}
	\caption{\label{fig:rdout-chain} Simpliﬁed block diagram of the SMX-based CBM
readout chain.
	}
\end{figure}

The \textcolor{black}{Compressed Baryonic Matter (CBM)} experiment in Darmstadt is being prepared. The production of
components for the readout chain is ongoing.
The key element for the readout of \textcolor{black}{Silicon Tracking System (STS) Muon Chamber (MUCH)} detectors is the
dedicated STS-MUCH-XYTER\footnote{\textcolor{black}{The full expansion of the STS-MUCH-XYTER is ``Silicon Tracking System - Muon Chamber - X-Y - Time -Energy Read-out''}} 2.2 (in short, SMX) ASIC developed by AGH
in Kraków~\cite{kasinski_sts-xyter_2014,maragoto_rodriguez_readout_2018}.
The SMX chip receives the pulses from the detector, digitizes and transmits them. 
The SMX ASICs are mounted on front-end boards (FEBs) in various
conﬁgurations depending on the detector system and expected hit
rates~\cite{romaniuk_gbt-based_2015}. %
The readout is done with GBTX-based~\cite{leitao_test_2015} Readout Boards (CROBs) further
connected to the Common Readout Interface (CRI)~\cite{zabolotny_cri_2018} data concentration backend. %
The communication between the CRI and CROBs is provided via so-called GBT Links (optical links with
4.8~Gb/s transmission speed) and between the CROBs and SMXes via AC-coupled differential SLVS links,
organized in so called E-Links.
The block diagram of the whole readout chain is shown in Figure~\ref{fig:rdout-chain}.

\section{Motivation for tester development}
Testing all produced ASICs, assembled \textcolor{black}{Front End Boards (FEBs)}, and full modules is the
essential quality veriﬁcation step in the production process~\cite{dogan_quality_2020,romaniuk_test_2016}.
\textcolor{black}{The whole testing process will include testing of approx. 40000 ASICs on a wafer during production, then approx. 2600 FEBs (tested multiple times during the assembly, e.g., after the ASIC wire bonding and after the micro-cable bonding), and finally the full module (consisting of a double-sided silicon strip sensor, connected by micro-cables to two FEBs, each with 8 SMX chips).
The production throughput will be approximately one module per day per production line. The manufacturing process is expected to be completed in two years.}
The SMX contains complex analog~\cite{kleczek_analog_2017}
 and digital~\cite{kasinski_back-end_2016} functionality controlled via a rich set of registers
accessible with a dedicated Hit Control Transfer Synchronous Protocol (HCTSP)~\cite{kasinski_protocol_2016}.
The same protocol is also used to send the received hit data to the data acquisition system.

The complete test procedure is relatively complex~\cite{romaniuk_test_2016} and requires
a possibility to control the tested ASICs from the software, receive and analyze
the collected hit data.

The simplest approach would be to use for tests a part of the full prototype readout chain. Unfortunately, both the GBTX chip and the CRI boards are subject to export restrictions and cannot be used in all countries developing or manufacturing SMX-based systems. The cost of such a solution would also be too high. 
The high cost and limited availability of components have also blocked the possibility of using a tester prepared to develop the digital part of SMX and communication protocol~\cite{zabolotny_design_2017}
 based on the \textcolor{black}{Kintex-based versatile AFCK board~\cite{url-afck} used as a readout backend in the prototype chain}.

As a hardware platform offering good availability and reasonable production cost, 
the \textcolor{black}{GBTX emulator (GBTxEMU)}~\cite{zabolotny_gbtx_2021} board was selected.

The GBTxEMU board was developed as a cheap and widely available
replacement for GBTX-based CROBs. 
It was developed for emulating the GBTX ASIC or the whole CROB board but may also be adapted
	for operation as an autonomous tester.
 It provides broad availability of
tester hardware for production in various production sites and additionally allows full
system tests in collaborating institutions worldwide independently of the
GBTX-based readout. It can be used either as an emulator of the GBTX
 readout board alone or as a full standalone backend system for frontend ASIC control.

	\section{GBTxEMU-based tester hardware}
The GBTxEMU board is based on a standard commercial Artix-7 board (TE-0712, Trenz Electronics GmbH).
The board is supplemented with a dedicated motherboard providing the application specific connectors and communication interfaces.
Additionally, it is equipped with a jitter cleaning device (SiLabs Si5344), which allows for clock recovery and system-wide synchronization if used in GBTX emulating mode.

The AC-coupled LVDS links are used as E-Links connected to SLVS  pins in SMX.

\begin{figure}[htbp]
	\begin{minipage}[h]{0.45\linewidth}
		\centering %
		\includegraphics[width=\linewidth]{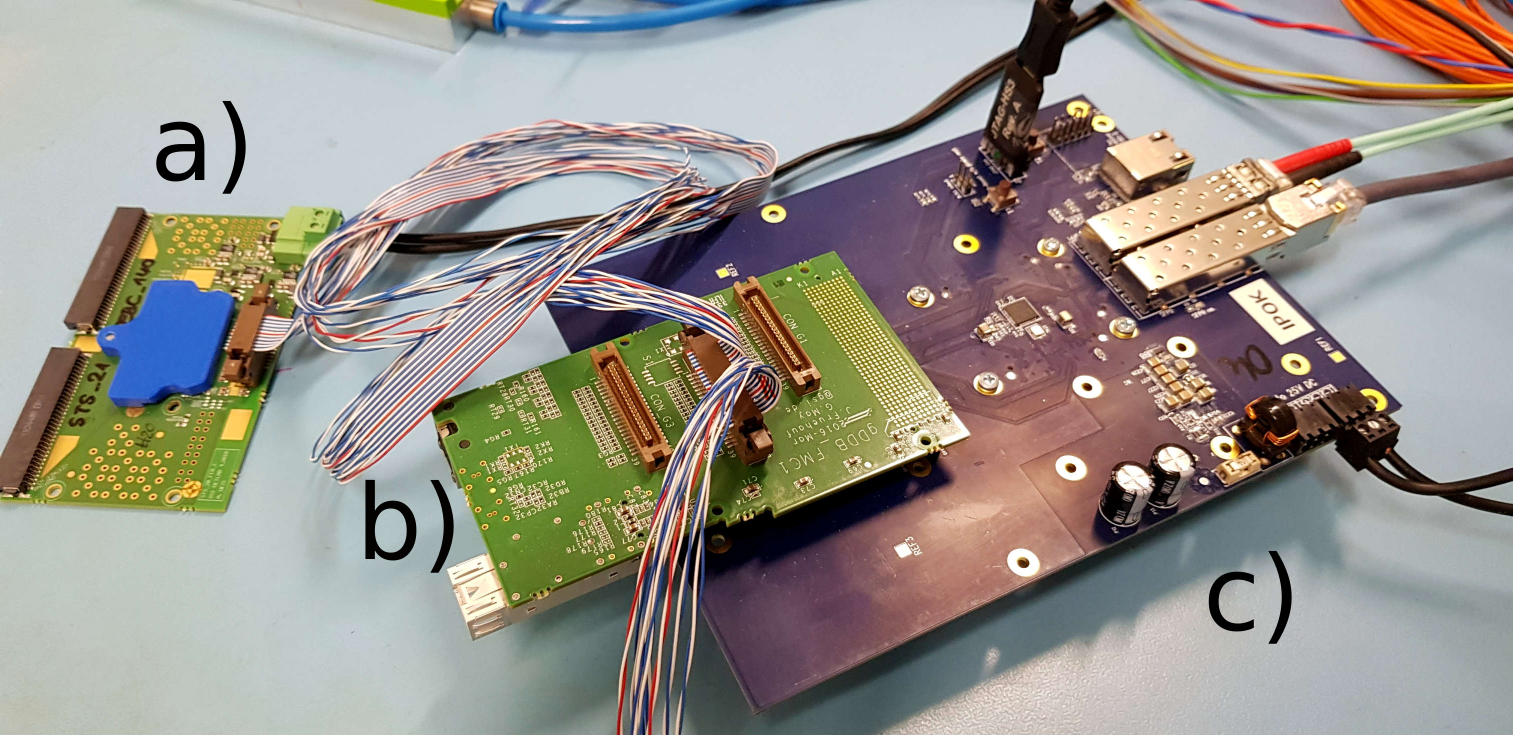}
	\end{minipage}
	\hspace{0.09\linewidth}
	\begin{minipage}[h]{0.45\linewidth}
		\centering %
		\includegraphics[width=\linewidth]{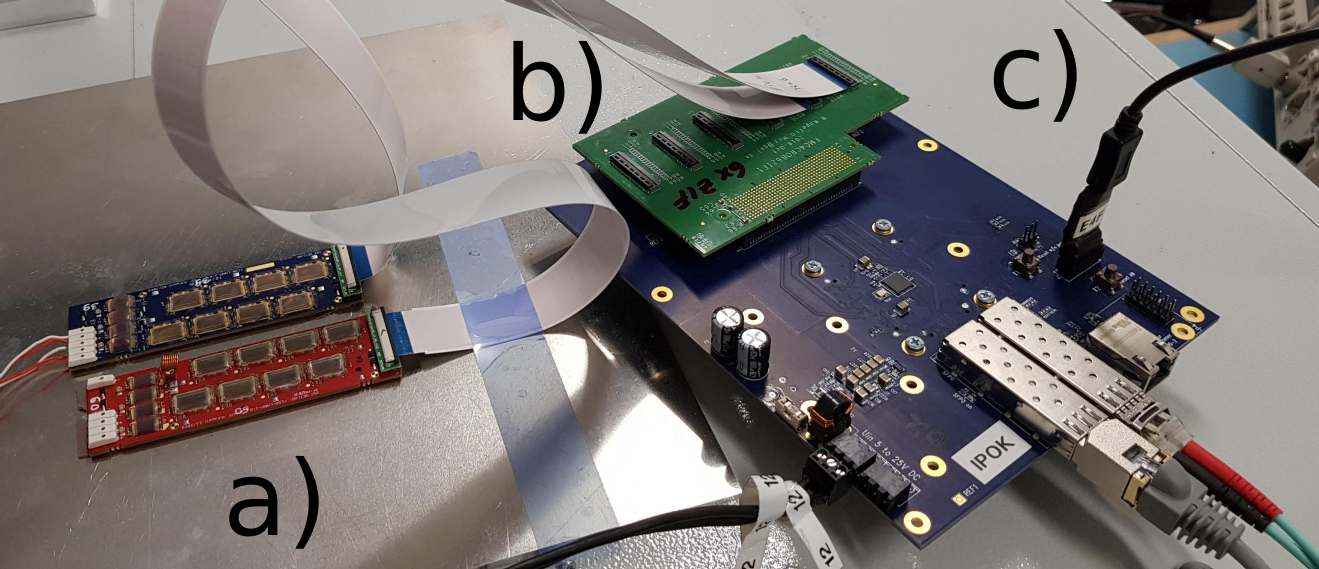}
	\end{minipage}
	\caption{\label{fig:fmcs} On the left - tester \textcolor{black}{(c)} with the FMC board \textcolor{black}{(b)} enabling connection of FEB-C
	boards or pogo-pin tester (one FEB-C with a single SMX is connected \textcolor{black}{(a)}). 
	On the right - tester \textcolor{black}{(c)} with the FMC board \textcolor{black}{(b)} enabling connection of FEB8 boards (two such boards, each with 8 SMXes are connected \textcolor{black}{(a)}).
	}
\end{figure}

A possibility to communicate with the SMX ASICs on various FEBs~\cite{romaniuk_gbt-based_2015}, with 
different numbers of connected ASICs and readout E-Links, has been provided using
dedicated VITA 57.1 FMC adapters (see Figure~\ref{fig:fmcs}). 
Special software for automated handling of connection lists has been created~\cite{kruszewski_safe_2021} to minimize
the risk of error when working with multiple interconnection schemes.

\section{GBTxEMU-based tester firmware}
The development of the tester firmware was founded on the experience gained during the development of \textcolor{black}{the protocol test environment~\cite{zabolotny_design_2017}, Data Processing Boards (DPBs) and
the CRI-based readout chains~\cite{cbm-prog-rep-2020-WUT}, %
and the GBTxEMU board~\cite{zabolotny_gbtx_2021}.}

\begin{figure}[htbp]
	\centering %
	\includegraphics[width=0.8\linewidth]{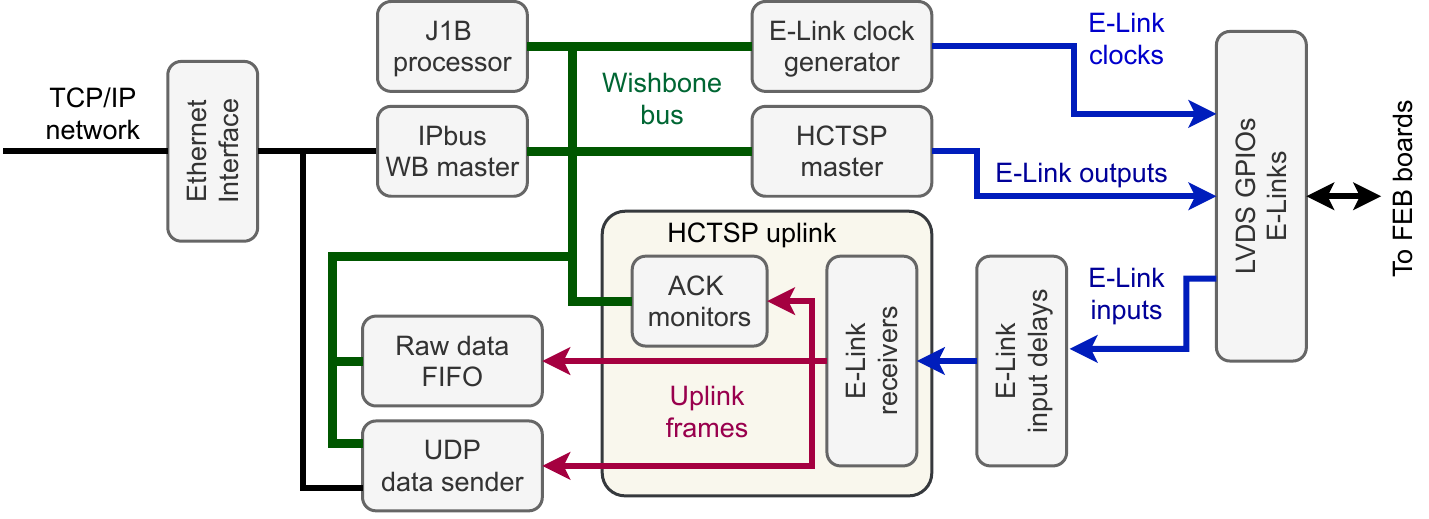}
	\caption{\label{fig:blk-diag} The block diagram of the tester firmware.
	}
\end{figure}

The tester firmware contains the HCTSP master block that generates the commands sent as HCTSP downlink frames via a single E-Link or a group of E-Links.
The E-Link inputs are connected to the HCTSP uplink block. The E-Link receiver blocks extract the uplink frames and deliver them to the  ACK monitors (that detect the command acknowledgments and responses), the raw data FIFO 
and the UDP data sender.
The block diagram of the tester firmware is shown in Figure~\ref{fig:blk-diag}.

\subsection{E-Link emulation}
The important features of the GBTX chip are the adjustment of the downlink clock phase
and adjustable delay of the E-Link input data. They enable adaptation to different lengths of the 
FEB-GBTX cable and to different delays between the clock and data lines.
The tester implements those functionalities using the \textcolor{black}{FPGA mixed-mode clock manager (MMCM) blocks~\cite{url-mmcm}}. The downlink clock frequency 
may be set to 40, 80, or 160 MHz. The clock phase may be adjusted with a resolution of 1/1600~MHz/8 = 78~ps 
(slightly worse than 48.8~ps offered by GBTX).
The E-Link input data delay may be adjusted with the IDELAYE2 blocks providing
the same resolution of 78 ps with a 200~MHz reference clock.

\section{The tester software}
The tester is controlled via the IPbus~\cite{larrea_ipbus:_2015} interface over 1 Gb/s Ethernet.
With the accompanying software written in Python, the tester supports
full testing of the E-Link communication. It also provides access to all
internal registers of the SMX chips, enabling complete functional tests.
The received ASIC hit data may be stored in an internal FIFO accessible
via IPbus, enabling the direct collection of a limited amount of hit data
at a high hit rate. For longer tests at a limited hit rate, the possibility to
transmit the hit data from selected uplinks in UDP packets has been
implemented.
\section{Tester in GBTX-emulation mode}
\textcolor{black}{The tester is based on the GBTxEMU hardware platform. Therefore,
after loading the GBTxEMU firmware, it may be used as an emulator of the CROB board
(see Figure~\ref{fig:rdout-chain})}. That
allows the sites equipped with the prototype versions of the CBM
readout chain to connect the tester via an emulated GBT optical link and
perform long-time tests at a high rate.

\section{Conclusions}
The tester is a versatile tool enabling testing of the SMX ASICs at the
manufacturing site as well as mounted FEBs and assembled detector
modules - both during manufacturing and after delivery, before mounting
in the detector. \textcolor{black}{Approximately 20 testers will be used in four production sites.
This number can be significantly increased if the parallel long-term testing of the assembled modules appears necessary.}
The FPGA-based design offers high ﬂexibility. The Python-based software may also be easily modified and adapted to the user's particular needs, including interfacing with the higher-level operation and quality verification software.

\acknowledgments

The work has been partially supported by GSI, 
and partially by the statutory funds of the Institute of Electronic Systems.

\bibliography{tester}
\bibliographystyle{JHEP}

\end{document}

%% file: wz_commands.tex
\usepackage[frozencache=true,cachedir=minted-cache]{minted}
\usepackage{multicol}
\usepackage{cprotect}

\newcommand{\wzurl}[1]{\href{#1}{\url{#1}}}